\documentclass[preprint,aps]{revtex4}
\usepackage{graphicx}

\begin{document}

\title{Antiferromagnetism and superfluidity
of a dipolar Fermi gas in a 2D optical lattice}
\author{Bo Liu, Lan Yin}
\email{yinlan@pku.edu.cn} \affiliation{School of Physics, Peking
University, Beijing 100871, China}
%\date{\today}

\begin{abstract}
In a dipolar Fermi gas, the dipole-dipole interaction between
fermions can be turned into a dipolar Ising interaction between
pseudospins in the presence of an AC electric field.
When trapped in a 2D optical lattice, such a dipolar Fermi gas has
a very rich phase diagram at zero temperature, due to the competition
between antiferromagnetism and superfluidity. At half filling,
the antiferromagnetic state is the favored ground state.  The
superfluid state appears as the ground state at a smaller filling
factor.  In between there is a phase-separated region.  The order
parameter of the superfluid state
can display different symmetries depending on the filling factor and
interaction strength, including d-wave ($d$), extended s-wave ($xs$), or
their linear combination ($xs+i\times d$).  The implication for the current experiment is discussed.
\end{abstract}

\maketitle

\section{Introduction}
The successful creation of $^{40}$K$^{87}$Rb polar molecules
with high space density \cite{Ni} has sparked a lot of research
interests because of its potential applications in various areas \cite{Carr}.
An ultracold gas of polar molecules can provide a platform to study
strongly-correlated systems \cite{Baranov}, due to the confinement
flexibility and the control over effective interactions \cite{Zoller}.
The dipole-dipole interaction can be turned into different effective
interactions in the presence of  a DC or AC electric field.
When the dipole direction is fixed by a DC electric field, a p-wave
superfluid state with the dominant $p_z$ symmetry may appear in a 3D dipolar
Fermi gas \cite{You}.  When an AC electric field is applied to couple
different rotational levels of molecules, a $p_x+ip_y$ superfluid state
\cite{Cooper} and a ferro-electric state
\cite{Wang} may show up in quasi-2D dipolar Fermi gases.  Creating
the antiferromagnetic state and exotic superfluid states such as p-wave and d-wave states
has been a focused goal in the research with optical lattices.
In this work we predict that some of these phases can be created in polar
molecules trapped in a 2D
square optical lattice in the presence of a resonant AC electric field.

For the $^{40}$K$^{87}$Rb molecule, rotational
excitation energies are much smaller than electronic and vibrational
excitation energies.  In the presence of the resonant AC electric field
for the first two rotational states, electronic and vibrational excitations
can be ignored, and the rotational state of the dipole along or against the
electric-field direction can represented by pseudospin-up or pseudospin-down state.
The dipole-dipole interaction is thus effectively a dipolar Ising interaction
between pseudospins.  The strength of this Ising interaction can be much
larger than the super-exchange interaction in a unpolarized two-component Fermi gas
trapped in the same optical lattice. In this aspect it is easier to create
antiferromagnetic and superfluid states with exotic symmetries in the dipolar Fermi gas.

This work is organized as follows.  First we derive the effective Hamiltonian
for this system.  Then the antiferromagnetic state and
superfluid states with different symmetries are
studied in the mean-field approximation.   The ground state is determined by comparing
energies of these states.   We find that the antiferromagnetic state is
the ground state at half filling.  When the filling factor
is reduced to certain values, the superfluid state takes over.  Depending
on the parameters, different superfluid states, such as d-wave ($d$),
extend s-wave ($xs$), and  the mixed-symmetry ($xs+i\times d$) states,
can appear.  Phase separation occurs between superfluid and antiferromagnetic
phases.  Based on these results, the phase diagram at zero temperature is obtained and
its experimental implication is discussed.
\section{Effective Model}
  The Hamiltonian of a single $^{40}$K$^{87}$Rb molecule trapped in a two-dimensional
optical lattice in the presence of an AC electric field is given by
\begin{equation}
H_0={p^2\over 2m} +H_{opt}+H_{rot},
\end{equation}
where the first r.-h.-s. term is the molecule kinetic energy for the center-of-mass
motion, and the second r.-h.-s. term is the potential energy in the optical lattice.
The rotational energy of the molecule in the presence of the
AC electric field is given by $H_{rot}=BL^2-{\bf d}\cdot {\bf E}_{AC}(t)$,
where ${\bf L}$ is the angular momentum operator, $B$ is the rotational constant,
{\bf d} is dipole operator, and ${\bf E}_{AC}(t)$ is the AC electric field.  The
eigenstate of the angular momentum is given by $|j,m\rangle$, where $j$ and
$m$ are quantum numbers of the total angular momentum and its z-component.  In this
work we consider a linearly polarized AC field in the $\hat{z}$ direction with
amplitude $E_{AC}$, ${\bf E}_{AC}(t)=E_{AC}\cos(\omega t)\hat{z}$.  The frequency
${\omega}$ is set to couple rotational states $|0,0\rangle$ and $|1,0\rangle$,
$\langle0,0|{\bf d}\cdot{\bf E}_{AC}(t)|1,0\rangle=dE_{AC}\cos(\omega t)/\sqrt{3}$,
where $d$ is the dipole moment.
Under the rotating wave approximation \cite{Wang} in the basis of $|0,0\rangle$ and
$\exp(-i\omega t)|1,0\rangle$, the molecule rotational energy is given by
$$H_{rot}=\left(\begin{array}{cccc}
0& \Omega \\ \Omega& 2B-\hbar\omega
\end{array}\right),$$
where $\Omega=dE_{AC}/\sqrt{3}$.  In the following we concentrate on the resonance
limit $\omega\rightarrow 2B/\hbar$ with small electric field, where the two rotational
states are nearly degenerate.

The dipole-dipole interaction is given by
\begin{equation}
V_{dd}({\bf r}_i-{\bf r}_j)= \frac{{\bf d}_i\cdot{\bf d}_j-3({\bf d}_i\cdot {\bf e}_{ij})
({\bf d}_j\cdot {\bf e}_{ij})}{\mid {\bf r}_i-{\bf r}_j \mid ^3 },
\end{equation}
where ${\bf d}_i$ is a dipole at ${\bf r}_i$, and ${\bf e}_{ij}$ is the direction of
${\bf r}_i-{\bf r}_j$.  In the presence of an resonant AC electric field,
only rotational states $|0,0\rangle$ and $|1,0\rangle$ are important.
The matrix elements of the dipole operator in this subspace are given by
$\langle0,0|{\bf d}|0,0\rangle=\langle1,0|{\bf d}|1,0\rangle=
\langle0,0|{\bf d}_{x,y}|1,0\rangle=0$, and $\langle0,0|d_z|1,0\rangle=d/\sqrt{3}$.
For this two-level system, we can define the pseudospin states
$|\uparrow\rangle  \equiv(|0,0\rangle+|1,0\rangle)/\sqrt{2}$ and
$|\downarrow\rangle \equiv(|0,0\rangle-|1,0\rangle)/\sqrt{2}$, so that the dipole matrix elements
are $\langle \uparrow |{\bf d}|\downarrow\rangle=0$,
$\langle \uparrow |{\bf d}|\uparrow\rangle=d\hat{z}/\sqrt{3}$, and
$\langle \downarrow |{\bf d}|\downarrow\rangle=-d\hat{z}/\sqrt{3}$.  In this new basis, the
interaction between two dipoles
can be conveniently written as a dipolar Ising interaction between pseudospins,
\begin{equation}
V_{dd}({\bf r}_i-{\bf r}_j)= 4d^2\frac{1-3\cos^2\theta_{ij}}
{3\mid \bf{r}_i-\bf{r}_j \mid ^3}s_{iz}s_{jz}
\end{equation}
where $s_{iz}$ is the $\hat{z}$ component of the pseudospin ${\bf s}_i$, and $\theta_{ij}$
is the angle between ${\bf e}_{ij}$
and the $\hat{z}$ direction. For this system, the optical lattice is set in the plane
perpendicular to the electric field, $\theta_{ij}=\pi/2$.

For a dipolar Fermi gas trapped in a two-dimensional optical lattice in the presence of a
resonant AC electric field, the effective Hamiltonian is therefore given by
\begin{equation}
H=-\sum_{\langle i,j \rangle,\sigma}t(c_{i\sigma}^{\dagger}c_{j\sigma}+c_{j\sigma}^{\dagger}c_{i\sigma})
+\frac{1}{2}\sum_{i\neq j}J_{ij}s_{iz}s_{jz}+U\sum_{i}n_{i\uparrow}n_{i\downarrow},
\label{hamiltonian 1}
\end{equation}
where $c_{i\sigma}$ is the fermion annihilation operator for site $i$ and spin $\sigma$,
$n_{i\sigma}=c^\dagger_{i\sigma}c_{i\sigma}$,
and the spin coupling constant between fermions at positions ${\bf r}_i$ and ${\bf r}_j$
is given by $J_{ij}=4d^2/(3|{\bf r}_j-{\bf r}_i|^3)$.  For simplicity we have assumed that
in the optical lattice the fermion can only hop between nearest neighbors with hopping amplitude $t$.
The onsite-interaction $U$ in Eq. (\ref{hamiltonian 1}) includes both the s-wave scattering between
different spin components and the onsite dipole interaction, and it can be tuned either
by using Feshbach resonance or by changing the vertical confinement \cite{Lahaye, Trefzger}.
Here we concentrate on the effect of the dipolar Ising interaction and set $U=0$ in the rest of this paper.

Since the dipolar Ising interaction is a short-range interaction in two dimension
and the spin coupling constant is the largest between nearest-neighbor spins, the system
tends to lower its energy by placing fermions with opposite spins at nearest-neighbor sites,
which can lead to two different types of ground states.  The first one is the antiferromagnetic state in
which spins align oppositely at nearest neighbors.  The other type is the superfluid
state with exotic symmetries where a fermion attracts another fermion with the opposite spin
at nearest neighbors to form a Cooper pair.   Under the current experimental condition \cite{Ni}, the
strength of the dipolar Ising interaction between nearest neighbors is approximately
$J\equiv4d^2/(3a^3)=300\hbar Hz$ for the singlet rovibrational state, the hopping amplitude $t$ is
close to $100\hbar$ $Hz$ \cite{Wall} for a typical optical lattice with the strength ${{V_{0}}/ {E_R}}=10$,
where $V_{L}(x)=V_{0}\sin(k_xx)$, $E_R=\hbar^2 k_x^2/2 m$, and the lattice constant is $a=1\mu m$.
In contrast, the super-exchange interaction of an ordinary Fermi gas is usually much smaller than
the hopping amplitude $t$, which makes it more difficult to produce ordered states.  The dipolar
Fermi gas has another advantage that the spin fluctuation
in Ising systems is much less than that in Heisenberg systems, which is helpful to stabilize ordered
states.  For these reasons, the superfluid and antiferromagnetic states may be easier to be created in the
dipolar Fermi gas than in an ordinary Fermi gas.  In the following we focus on the properties of these
ordered states at zero temperature.

\section{Antiferromagnetic state}
To study the antiferromagnetic state, we divide the square lattice into two sublattices A and B so that
nearest neighbors of one sublattice all belong to the other sublattice.  The antiferromagnetic order
is described by the finite average of the pseudospin, $\langle s_{iz} \rangle=m$
for sublattice A and $\langle s_{jz} \rangle=-m$ for sublattice B.
In the ${\bf k}$-space, the mean-field Hamiltonian for the antiferromagnetic state is given by
\begin{eqnarray}
H_{AF}&=&-{hm\over2}\sum_{{\bf k}}(a_{{\bf k}\uparrow}^{\dagger}a_{{\bf k}\uparrow}+b_{{\bf k}\downarrow}^{\dagger}b_{{\bf k}\downarrow}-
a_{{\bf k}\downarrow}^{\dagger}a_{{\bf k}\downarrow}-b_{{\bf k}\uparrow}^{\dagger}b_{{\bf k}\uparrow})\nonumber\\
&+&\sum_{{\bf k},\sigma}\varepsilon_{\bf k}(a_{{\bf k}\sigma}^{\dagger}b_{{\bf k}\sigma}+b_{{\bf k}\sigma}^{\dagger}a_{{\bf k}\sigma})
+\frac{h m^2 N}{2},\label{AFH}
\end{eqnarray}
where $N$ is the total number of lattice sites, the ${\bf k}$-summation is over the first Brillouin zone of a sublattice,
$a_{{\bf k}\sigma}$ and $b_{{\bf k}\sigma}$ are fermion annihilation operators for sublattice A and B,
$\varepsilon_{\bf k}=-2t(\cos k_xa+\cos k_ya)$, and $a$ is the lattice constant.  The constant
$h$ can be regarded as the strength of the effective
antiferromagnetic interaction, $h \equiv- \sum_{i\neq 0}(-1)^{i}J_{0i}=2.646J$.

The mean-field Hamiltonian in Eq. (\ref{AFH}) can be diagonalized by a standard canonical transformation,
\begin{equation}
H_{AF}=\sum_{{\bf k},\sigma}(\varepsilon_{\bf k}^{-}\alpha_{{\bf k}\sigma}^{\dagger}\alpha_{{\bf k}\sigma}+\varepsilon_{\bf k}^{+}\beta_{{\bf k}\sigma}^{\dagger}\beta_{{\bf k}\sigma})+\frac{h m^2 N}{2},
\label{AFMF 2}
\end{equation}
where the quasi-particle operators are given by $\alpha_{{\bf k}\uparrow}=u_{\bf k}a_{{\bf k}\uparrow}+v_{\bf k}b_{{\bf k}\uparrow}$,
$\alpha_{{\bf k}\downarrow}=u_{\bf k}b_{{\bf k}\downarrow}+v_{\bf k}a_{{\bf k}\downarrow}$,
$\beta_{{\bf k}\uparrow}=u_{\bf k}b_{{\bf k}\uparrow}-v_{\bf k}a_{{\bf k}\uparrow}$,
$\beta_{{\bf k}\downarrow}=u_{\bf k}a_{{\bf k}\downarrow}-v_{\bf k}b_{{\bf k}\downarrow}$, and the transformation coefficients are given by
$u_{\bf k}^2=1-v_{\bf k}^2=[1+h m/(2\varepsilon^+_{\bf k})]/2$.  The quasi-particles form two bands with energies given by
 \begin{equation}
\varepsilon_{\bf k}^{\pm}=\pm\sqrt{\varepsilon^2_{\bf k}+(\frac{h m}{2})^2}.
\end{equation}
The magnetic moment $m$ can be determined self-consistently from the equation
\begin{equation}
m=\frac{1}{N}\sum_{\bf k}\langle a_{{\bf k}\uparrow}^{\dagger}a_{{\bf k}\uparrow}-a_{{\bf k}\downarrow}^{\dagger}a_{{\bf k}\downarrow}\rangle,
\end{equation}
which leads to the equation
\begin{equation}
{1 \over h}=\frac{1}{N}\sum_{\bf k}{1 \over 2\varepsilon^+_{\bf k}}[\theta(\mu-\varepsilon_{\bf k}^{-})-\theta(\mu-\varepsilon_{\bf k}^{+})],
\label{AFSC}
\end{equation}
where $\mu$ is the chemical potential.  Eq.~(\ref{AFSC}) is equivalent to the energy-extreme condition  $\partial E / \partial m=0$,
and can be solved together with the equation for the fermion number per site
\begin{equation}
n=\frac{2}{N}\sum_{\bf k}[\theta(\mu-\varepsilon_{\bf k}^{-})+\theta(\mu-\varepsilon_{\bf k}^{+})].
\label{AFNE}
\end{equation}
The total energy per site is given by
\begin{equation}
E_{AF}=\frac{2}{N}\sum_{\bf k}[\varepsilon_{\bf k}^{-}\theta(\mu-\varepsilon_{\bf k}^{-})+\varepsilon_{\bf k}^{+}\theta(\mu-\varepsilon_{\bf k}^{+})]+\frac{h m^2}{2}.
\label{AFEN}
\end{equation}

\begin{figure}[t]
\begin{center}
\includegraphics[width=8cm]{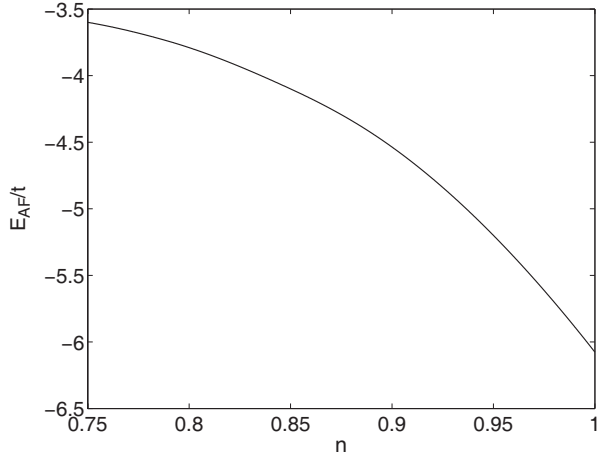}
\caption{The energy of the antiferromagnetic state as a function of the filling factor $n$ at $J/t=3$.  The downward curvature shows
that the compressibility is negative and antiferromagnetic state is unstable beyond half filling. } \label{figure1}
\end{center}
\end{figure}

At half filling $n=1$, the r.-h.-s. of Eq. (\ref{AFSC}) is infraredly divergent at $m=0$, indicating that
the antiferromagnetic order always exists for any finite dipolar Ising interaction.
The antiferromagnetic solution can also be obtained at other filling factors as long as the following condition is satisfied,
\begin{equation}
{1 \over h}\leq\frac{1}{N}\sum_{\bf k}{1 \over 2\varepsilon_{\bf k}}[\theta(\mu+|\varepsilon_{\bf k}|)-\theta(\mu-|\varepsilon_{\bf k}|)].
\end{equation}
However, as shown in Fig. \ref{figure1}, the compressibility of the antiferromagnetic state turns negative,
indicating that the antiferromagnetic state is dynamically unstable beyond half filling.

\section{Superfluid states}
   At the superfluid state, two fermions with opposite spins form a pair.
In the Hamiltonian in Eq. (\ref{hamiltonian 1}), this pairing interaction is given by
$-\sum_{i\neq j}J_{ij}c^{\dagger}_{i\uparrow}c_{i\uparrow}c^{\dagger}_{j\downarrow}c_{j\downarrow}/4$.
In the ${\bf k}$-space, the mean-field Hamiltonian for the superfluid state is given by
\begin{equation}\label{MFSFHM 1}
H_{SF}=\sum_{{\bf k}\sigma}(\varepsilon_{\bf k}-\mu)c_{{\bf k}\sigma}^{\dagger}c_{{\bf k}\sigma}
+\sum_{{\bf k}}(\Delta^*_{{\bf k}}c_{-{\bf k}\downarrow}c_{{\bf k}\uparrow}+
\Delta_{{\bf k}}c^{\dagger}_{{\bf k}\uparrow}c^{\dagger}_{-{\bf k}\downarrow}-\Delta_{{\bf k}}g^*_{{\bf k}}),
\end{equation}
where the pairing gap is given by $\Delta_{\bf k}=\sum_{{\bf k}^{\prime}}f_{{\bf k}{\bf k}^{\prime}}g_{{\bf k}'}$,
$g_{\bf k}=\langle c_{-{\bf k}\downarrow}c_{{\bf k}\uparrow}\rangle$,
and $$f_{{\bf k}{\bf k}'}=-{1 \over 4N^2} \sum_{i\neq j}J_{ij}\exp[i({\bf k}-{\bf k}')\cdot ({\bf r}_i-{\bf r}_j)].$$

The mean-field Hamiltonian Eq. (\ref{MFSFHM 1}) can be diagonalized by Bogoliubov transformation
as in the BCS theory,
\begin{equation}
H_{SF}=N(E_{SF}-\mu n)+\sum_{{\bf k}\sigma}E_{\bf k}d^{\dagger}_{{\bf k}\sigma}d_{{\bf k}\sigma},
\label{MFSFHM 2}
\end{equation}
where $E_{\bf k}=\sqrt{(\varepsilon_{\bf k}-\mu)^2+|\Delta_{\bf k}|^2}$ is the quasiparticle energy,
$d_{{\bf k}\sigma}$ is the quasi-particle annihilation operator,
and $E_{SF}$ is the ground-state energy per site,
\begin{equation}\label{SFEN}
E_{SF}=\frac{1}{N}\sum_{\bf k}(\varepsilon_{\bf k}-\mu-E_{\bf k}+\frac{|\Delta_{\bf k}|^2}{2E_{\bf k}})+\mu n.
\end{equation}
The gap $\Delta_{\bf k}$ can be determined from the gap equation
\begin{equation}
\Delta_{\bf k}=-{1\over N}\sum_{{\bf k}^{\prime}}f_{{\bf kk}^{\prime}}\frac{\Delta_{{\bf k}^{\prime}}}{2E_{{\bf k}^{\prime}}}.
\label{gap}
\end{equation}
The fermion number per site is given by
\begin{equation}
n=1-\frac{1}{N}\sum_{\bf k}\frac{\varepsilon_{\bf k}-\mu}{E_{\bf k}}.
\label{density 1}
\end{equation}

\begin{figure}[t]
\begin{center}
\includegraphics[width=8cm]{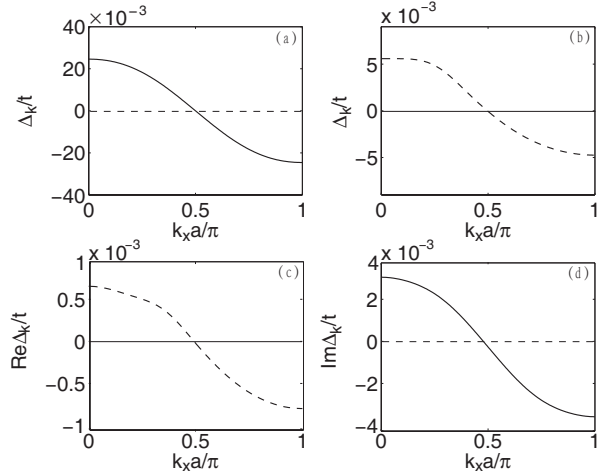}
\caption{The gap $\Delta_{\bf k}$ versus the $x$ component of the wavevector $k_x$.  The dashed and solid lines are for $\Delta_{\bf k}$ at $k_x=k_y$ and $k_x+k_y=\pi/a$ respectively.
(a) At $J/t=3$ and $n=0.656$, the gap $\Delta_{\bf k}$ shows the d-wave symmetry.  (b) At $J/t=3$ and $n=0.29$, the gap shows
the extended-s-wave symmetry.  At $J/t=3$ and $n=0.483$, (c) the real part of the gap shows the extended-s-wave symmetry, and (d) the imaginary part of the
gap shows d-wave symmetry.}  \label{figure2}
\end{center}
\end{figure}

We numerically solve the gap equation (\ref{gap}) together with Eq. (\ref{density 1}), and find that the gap $\Delta_{\bf k}$ can display different
symmetries depending on the parameters.  The d-wave superfluid state appears in the region with $0.5<n<1$ and $J<5t$.  As shown in Fig. \ref{figure2}(a),
the d-wave symmetry is characterized by the sign change in the gap at $k_x \pm k_y=0$.  The extended-s-wave ($xs$)
superfluid state may appear for $n<0.5$ with the sign change of the gap at $k_x \pm k_y=\pi/a$ as shown in \ref{figure2}(b).
Between the d-wave and extended s-wave phases,
a superfluid phase appears with a mixed $xs+i\times d$ symmetry in the order-parameter.  In this phase,
a global unitary transformation can be always applied so that the real part of the gap has the extended-s-wave symmetry, and
the imaginary part of the gap has the d-wave symmetry, as shown in Fig. \ref{figure2}(c) and \ref{figure2}(d).   Unlike the d-wave or extend-s-wave
state, this mixed-symmetry state breaks the time-reversal symmetry, and supports the spontaneous current \cite{SDM}.

\section{Zero-Temperature Phase Diagram}
\begin{figure}[t]
\begin{center}
\includegraphics[width=8cm]{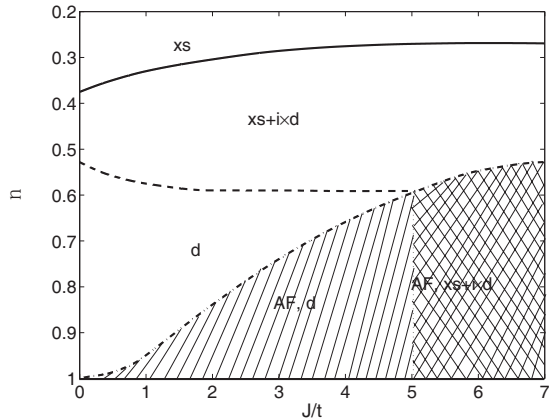}
\caption{Phase diagram at zero temperature. The antiferromagnetic phase is along the half-filling line.
The dashed-dotted line marks the boundary between the phase-separation region and the superfluid phase.
The dashed (solid) line is the phase boundary between the mixed-symmetry and the d-wave
(extend-s-wave) superfluid phases.} \label{figure3}
\end{center}
\end{figure}

  By comparing the energies of the antiferromagnetic and superfluid states, we obtain
the zero-temperature phase diagram of this system, as shown in Fig. \ref{figure3}.   At half filling, the antiferromagnetic
state is the ground state, but it is unstable beyond half filling.  When $J<5t$,
the d-wave superfluid state appears at a smaller filling factor.  Phase separation between the antiferromagnetic and
d-wave superfluid states takes place in between.  As the filling factor further decreases,
the d-wave superfluid state evolves into the mixed-symmetry ($xs+i\times d$) state.  As the filling factor continues to decrease,
this mixed-symmetry eventually evolves into the extended-s-wave superfluid state.

When $J>5t$, the above picture slightly changes.  The d-wave superfluid state no longer shows up.  As the filling
factor decreases from half filling, a superfluid state appears with the mixed-symmetry.  Phase separation
between the antiferromagnetic and mixed-symmetry superfluid states occurs in the middle.  As the filling factor further decreases,
this mixed-symmetry evolves into the extended-s-wave superfluid state.

In the current experimental situation\cite{Ni}, the strength of the dipolar Ising interaction is approximately $J=300\hbar Hz$
for the singlet rovibrational state, and the hopping amplitude is approximately $100\hbar Hz$ \cite{Wall} for a typical optical
lattice with lattice constant $a=1\mu m$.   When the filling factor changes from zero to half filling,
extended-s-wave, mixed-symmetry, and d-wave superfluid phases, the phase-separation region,
and the antiferromagnetic state at half filling can all be reached as shown in Fig. \ref{figure3}.  The
hopping amplitude $t$ can be tuned by increasing or decreasing the strength of the optical-lattice potential, and other regions
in the phase diagram can also be reached.

\section{Discussion and Conclusion}
So far we have considered only the case for vanishing onsite interaction $U=0$.  With finite $U$, the phase diagram can be more
complicated.  If $U>0$ and $U\leq J$,  the antiferromagnetic state is more energetically favorable, and we expect a wider region of the phase
separation between the antiferromagnetic and superfluid states in the phase diagram.  If $U\gg J$, for low
energy states there is effectively a constraint of no more than one fermion per site as discussed in the $t-J-V-W$ model \cite{Demler},
which can strongly affect the superfluid states.  If $U<0$, a pure s-wave superfluid state may appear depending on the parameters.
All these cases are worth to explore in future works.

Experimentally the onsite interaction leads to losses of the KRb molecules through chemical reaction, which pose difficulties
for creating the antiferromagnetic and superfluid phases proposed in this paper.  Progress has been made to suppress the reaction rate in a quasi-2D geometry \cite{de Miranda}.
In other polar molecules systems, such as KCs, NaK and NaRb, there are no reactive channels\cite{Zuchowski}, and the loss problem can be
avoided.  With more experimental progresses, it is hopeful for observing the phase diagram shown in Fig. \ref{figure3}.

In summary, we have shown that in the presence of a resonant AC electric field, the dipole-dipole interaction in a dipolar Fermi gas
can be turned to a dipolar Ising interaction between pseudospins.  When trapped in a square optical lattice, the dominant antiferromagnetic
interaction between nearest neighbors drives the system into different ordered states depending on parameters.  The antiferromagnetic state is the ground
state at half filling.  The superfluid state appears as the ground state at a smaller filling factor.  In the middle is the phase-separated
region.  Depending on the parameters, the superfluid order parameter displays different symmetries, such as the d-wave, extended-s-wave, and the mixed symmetry.
These states mentioned above may be detectable under experimental conditions.
\section{Acknowledgment}
We would like to thank T.-L. Ho for helpful
discussions. This work is supported by NSFC under Grant No 10974004.

\end{document}